\documentclass[conference]{IEEEtran}
\IEEEoverridecommandlockouts
\usepackage{cite}
\usepackage{amsmath,amssymb,amsfonts}
\usepackage{algorithmic}
\usepackage{graphicx}
\usepackage{textcomp}
\usepackage{hyperref}
\usepackage{xcolor}
\usepackage{multirow}
\usepackage{listings}
\def\BibTeX{{\rm B\kern-.05em{\sc i\kern-.025em b}\kern-.08em
    T\kern-.1667em\lower.7ex\hbox{E}\kern-.125emX}}
\begin{document}

\title{Boosting Distributed Training Performance of the Unpadded BERT Model
}

\author{\IEEEauthorblockN{
Jinle Zeng\IEEEauthorrefmark{2}, 
Min Li\IEEEauthorrefmark{2},
Zhihua Wu\IEEEauthorrefmark{2}
\IEEEcompsocitemizethanks{\IEEEcompsocthanksitem\IEEEauthorrefmark{2}Equal contribution}}
\IEEEauthorblockN{Jiaqi Liu, Yuang Liu, Dianhai Yu, and, Yanjun Ma}

\textit{Baidu Inc., China}
}

{}

\maketitle

\begin{abstract}
Pre-training models are an important tool in Natural Language Processing (NLP), while the BERT model is a classic pre-training model whose structure has been widely adopted by followers. It was even chosen as the reference model for the MLPerf training benchmark. The distributed training performance optimization of BERT models plays an important role in accelerating the solutions of most NLP tasks. BERT model often uses padding tensors as its inputs, leading to excessive redundant computations. Thus, removing these redundant computations is essential to improve the distributed training performance. 

This paper designs a new approach to train BERT models with variable-length inputs efficiently. Firstly, we propose a general structure for the variable-length BERT models, and accelerate the encoder layer via our grouped multi-stream FMHA (Fused Multi-Head Attention) method. Secondly, through data exchange, we address the unbalanced workload problem caused by the variable-length inputs, which overlaps highly with the training process. Finally, we optimize the overall performance of the BERT model, such as kernel fusion, and operator optimization. Our experimental results show that our highly optimized BERT model achieves state-of-the-art throughput and ranks first in MLPerf Training v2.0 within the same GPU configuration. The optimizations in this paper can be applied to more BERT-like models in our future works.
\end{abstract}


\section{Introduction}


In recent years, many pre-training models emerged in Nature Language Processing (NLP) based on the Transformer structure (Ashish Vaswani et al.) in 2017 \cite{vaswani2017attention}. Among all these pre-training models, BERT (Bidirectional Encoder Representations from Transformers), introduced by Jacob Devlin et al. in 2018 \cite{2018-bert}, is the cornerstone for many following models. The BERT model achieved state-of-the-art results on 11 different NLP tasks by the time of the publication. Ever since then, many ground-breaking models in NLP have been updated or modified on the basis of the BERT model, such as ERNIE (Yu Sun et al.) \cite{sun2019ernie}, RoBERTa (Yinhan Liu et al.) \cite{liu2019roberta}, XLNet (Zhilin Yang et al.) \cite{yang2019xlnet}, ELECTRA (Kevin Clark et al.) \cite{clark2020electra}. Outside of the NLP paradigm, the Transformer also becomes one of the most prevailing structures in computer vision (Scaling Vision Transformer by Xiaohua Zhai et al. \cite{zhai2022scaling}) and speech recognition (Conformer by Anmol Gulati et al. \cite{gulati2020conformer}). The parameter size of the BERT-like models is increasing. The BERT large model only contained 340M parameters at first. Only two years later, the parameter size of GPT-3 (Tom B. Brown et al.) \cite{brown2020language} reached 175B. With the increase of the parameter size for these models, the cost to train these models is ramping up, boosting demand for optimization of the distributed training performance of BERT-like models. 

Training samples for NLP models are usually of variable length. A common strategy for training BERT models is to use padded inputs  \cite{rajbhandari2020zero, rasley2020deepspeed,shoeybi2019megatron, lightseq-train, yan2021fastseq, turbotransformers}. The inputs are usually padded to the maximum sequence length, making them fixed shape tensors. It is easy to build a model with padding inputs, but the information from the padding tokens should be removed in order to obtain the correct loss function. In addition, padding tokens introduce redundancy in the computation of many operators in the network. To improve the distributed training performance of the BERT model, it is important to remove the unnecessary computations in the padding tokens.

However, there are a number of challenges in developing BERT models with variable length inputs. On the one hand, many computation-related issues requires careful consideration. We need to design the input storage with caution and select parts of the model to be computed by variable-length inputs. In the case of variable-length storage, many operators need to be further optimized for better performance, such as the attention layer and embedding operator. Due to the variation in sequence length for different samples, it is difficult to implement a generic kernel that performs well for all cases. On the other hand, variable-length inputs have the potential to cause unbalanced workload problems in distributed training, leading to severe communication overheads and unsatisfactory distributed training performance.

In this paper, we propose a BERT model using the variable-length inputs instead of padding, and we name it the unpadded BERT model. We optimize the distributed training performance of the unpadded BERT model through various methods. First, we design the way to build the unpadded BERT model, and introduce the grouped multi-stream FMHA (Fused Multi-Head Attention) layer to improve the performance of the BERT encoders. Second, we solve the unbalanced workload problem in distributed training by exchanging data across the devices. The data exchanging process fully overlaps with the training process after our optimization. Last, we make some improvements to the CUDA kernels, such as kernel fusion and operator optimization. Our experiments showed that our highly optimized BERT model is the fastest among models like NVIDIA MLPerf BERT, HazyResearch MLPerf BERT, Megatron-LM, and DeepSpeed. Our optimized version has been submitted to the MLPerf Training v2.0, and ranked first within the same GPU configurations.

\section{Related Works}

Recently, many works are attempting to accelerate the Transformer-based model. FastSeq (Yu Yan et al.) \cite{yan2021fastseq} proposed an optimization method for inference with Transformer structure, including generation attention caching, repeated n-grams pattern detecting and asynchronous generation. TurboTransformers (Jiarui Fang et al.) \cite{turbotransformers} also introduced an optimization method for the inference, including an efficient parallel algorithm, a new memory allocation algorithm and a new batch sampler. Although the FastSeq and TurboTransoformer can accelerate the Transformer-based model with many techniques, both of them focus on the inference stage instead of training stage. In other words, they did not optimize the performance of backward and optimizer computations.

DeepSpeed \cite{rajbhandari2020zero, rasley2020deepspeed} and Megatron-LM \cite{shoeybi2019megatron} also made some training performance optimizations by CUDA kernel fusions and sharding the parameters across all the workers. LightSeq (Xiaohui Wang et al.) \cite{lightseq-train} brought an end-to-end system containing kernel fusions (covering embedding module, encoder module and criteria module), a mixed-precision training method with high memory efficiency, and a method for memory management. With the LightSeq system, the Transformer structure can be trained efficiently in the  training stage and inference stages. All these works create redundant computations due to the padding inputs.

NVIDIA proposed the unpadded implementation of the BERT model for the first time on MLPerf Training v1.0 \cite{mlperf_benchmark}, and continued the same implementation on MLPerf Training v1.1 and v2.0 \cite{mlperf-2.0}. But there are still some performance issues waiting to be improved like more unpad modules, overlapping data preprocessing with training. This paper is inspired by the NVIDIA MLPerf works, and focuses on a deeper distributed training optimization of the unpadded BERT model.

{}

{}

\section{Background}
\label{sec-background}

This section will introduce some background knowledge for the architecture and training strategy of the BERT model. Then the challenges are addressed to optimize the distributed training performance of the unpadded BERT model.

\subsection{Overview of the BERT Model}

\begin{figure}[!htb]
\centering
\includegraphics[width=0.8\columnwidth]{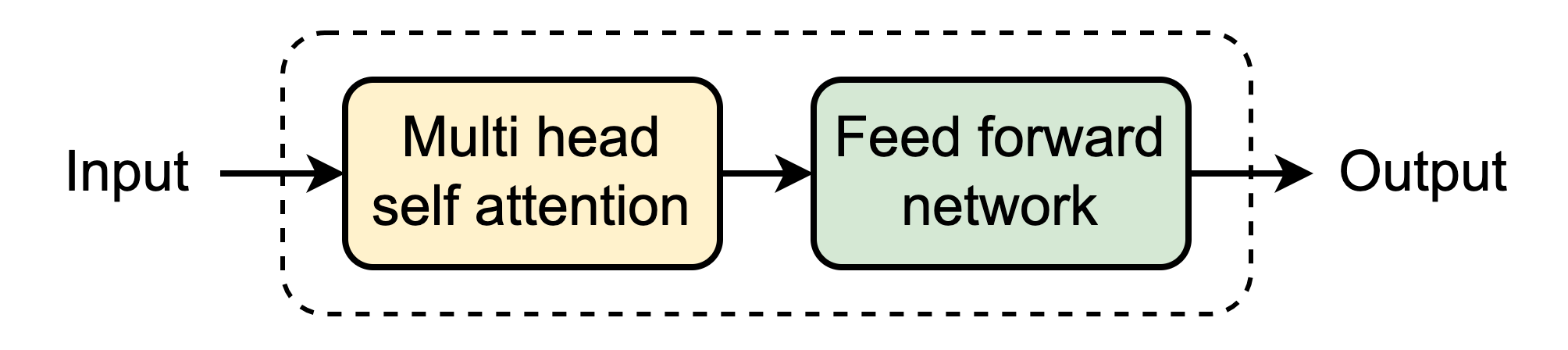}
\caption{Structure of the BERT encoder. }
\label{fig-encoder-layer}
\end{figure}

The core component of the BERT model is the stack of multiple encoder modules. Each encoder module contains one multi-head attention layer and one feed forward layer. The structure of the BERT encoder module is shown in Figure \ref{fig-encoder-layer}.

The workflow of a multi-head self-attention layer can be expressed as follows: 
\begin{equation}
Attention (Q, K, V) = softmax({\frac{QK^T}{\sqrt{d_k}} }) V,
\end{equation}
where $d_k$ represents the head dimension. 

\begin{figure}[!htb]
\centering
\includegraphics[width=0.8\columnwidth]{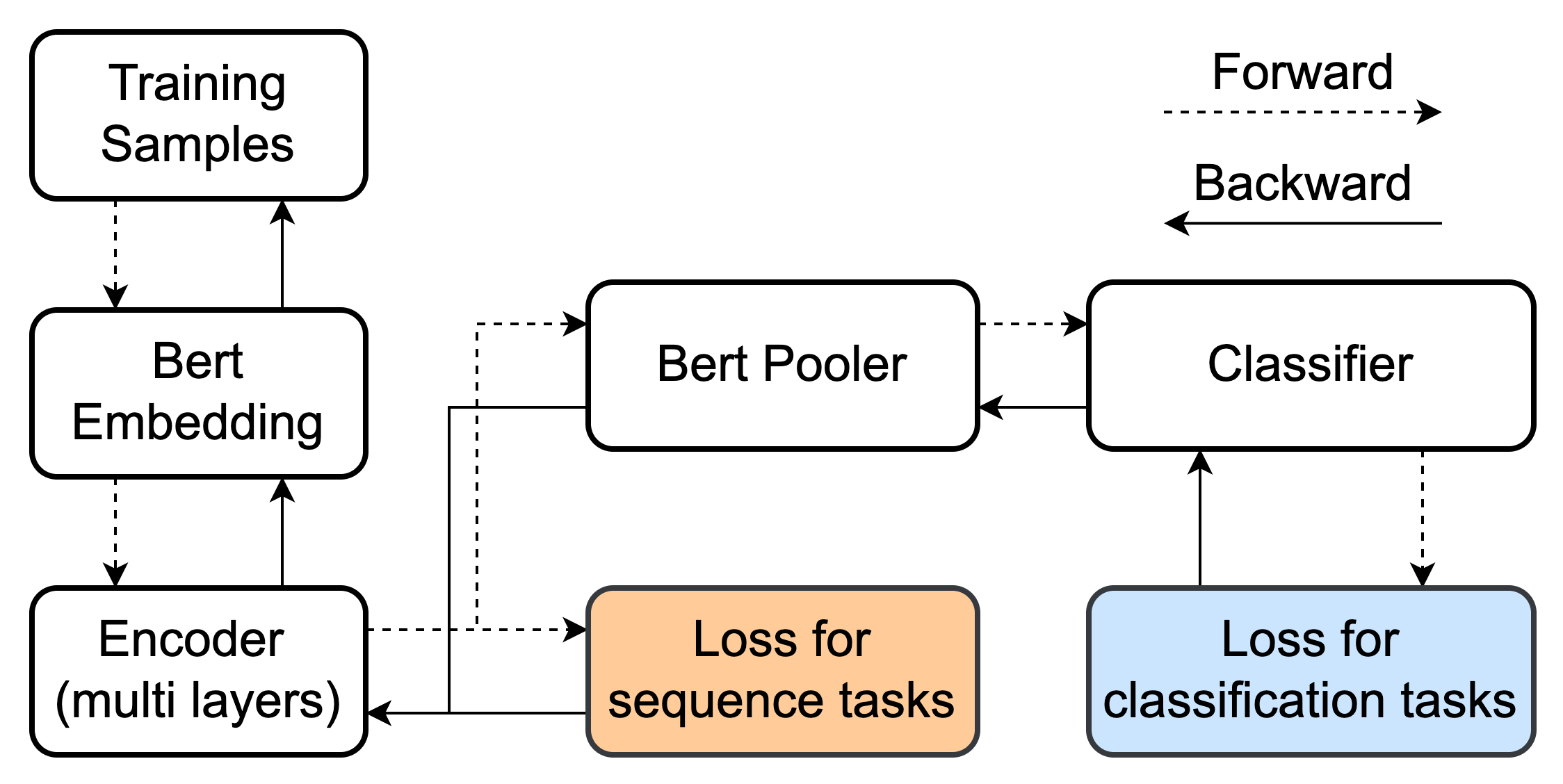}
\caption{The Training Process of the BERT Model.}
\label{bert-training}
\end{figure}

Besides the stack of multiple encoder layers, the pre-training BERT model also contains an embedding module, and a pooler module with a classifier for the classification tasks. The forward and backward flows of the BERT model training are presented in Figure \ref{bert-training}.

\subsection{Training Strategy of the BERT Model}


The training data of the language model is usually composed of variable-length sequences. Most existing training frameworks, such as DeepSpeed or Megatron-LM, use the padding method to ensure all input data in one batch are of the same length. For example, for the input data with a length less than \textit{max\_seq\_len}, several special tokens \textit{PAD} would be added to the end of the data to make the data length be \textit{max\_seq\_len}. After this step, we can change the variable-length sequences into fixed lengths. However, extra computations are introduced for these padding tokens.

\begin{figure}[!htb]
\centering
\includegraphics[width=0.9\columnwidth]{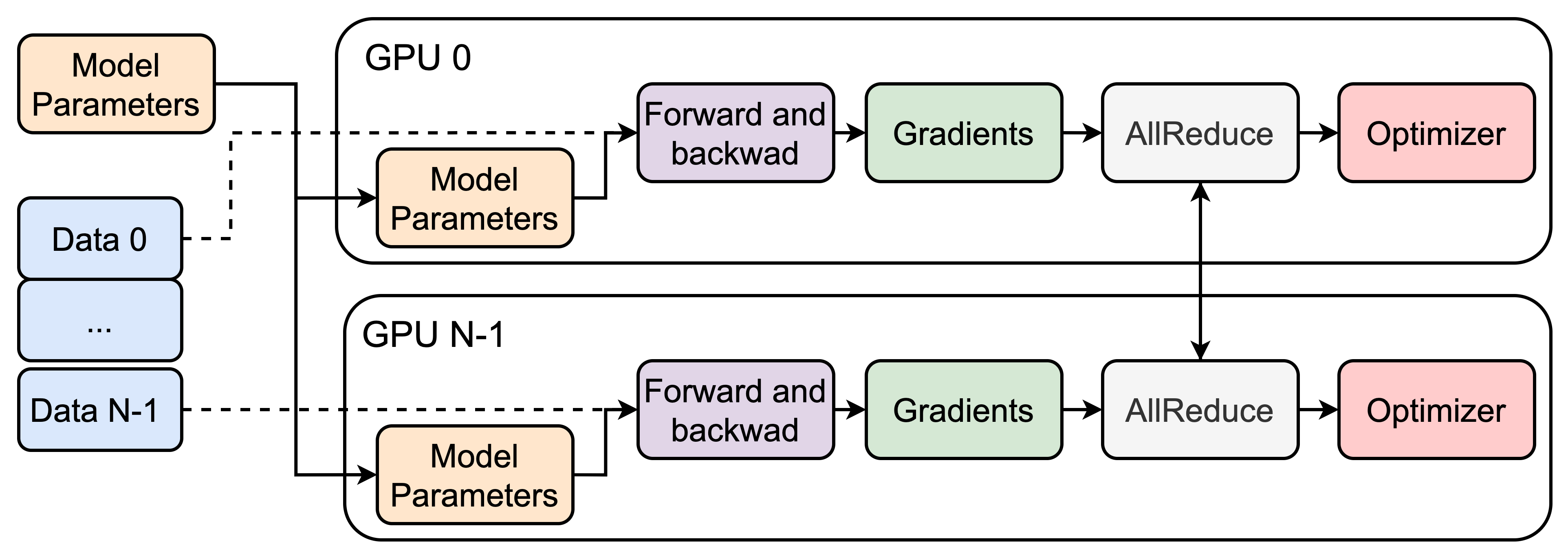}
\caption{The Data Parallel Training of BERT Model. }
\label{data-parallel}
\end{figure}

To speed up the training process and utilize the GPU resources, we use the data parallelism distributed strategy to train the BERT with multiple GPUs in this paper.
For data parallelism, each GPU contains the same model architecture and parameters, but the training data would be split into many parts. Each GPU will take one part of the data. After each GPU finishes the forward and backward parts, all GPUs would synchronize the gradients of each parameter by the all-reduce operator to carry out the optimizer part. The workflow of data parallelism is shown in Figure \ref{data-parallel}.

\subsection{Challenges of the Variable-Length BERT Optimization}

\subsubsection{Computational Optimizations for Variable-Length Inputs}

\begin{figure}[!htb]
\centering
\includegraphics[width=0.85\columnwidth]{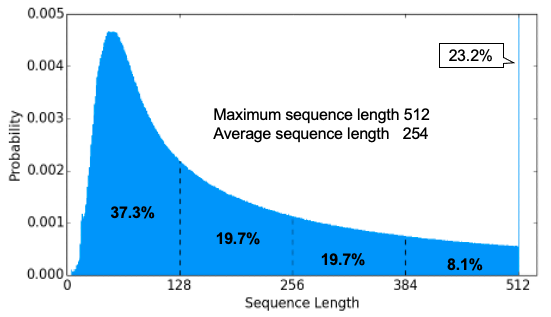}
\caption{The distribution of the sequence length for Wikipedia data set.}
\label{figure-varlen}
\end{figure}

As discussed above, the data sets of the language model are usually composed of a series of sequences with variable lengths. Most existing works rely on the padding method to deal with these inputs. In fact, the number of tokens for different sequences can vary drastically. We take Wikipedia data set as an example. As shown in Figure \ref{figure-varlen}, samples with maximum sequence length only account for 23.2\%. Therefore, the padding method may bring a large number of redundant computations, leading to unsatisfied performance. If we only compute on valid tokens, it is potential to bring more than 2x performance improvements.
Besides, the padding method would occupy more memory space, thus putting more pressure on the system memory. It is urgent to develop methods to support the unpad computing to achieve higher performance. However, supporting computing for variable-length inputs is challenging. The first challenge is that the input storage needs careful design, and deep analysis to decide which part of the model can perform the unpad computing. The second challenge is that many operators for the unpad computing require optimization to boost the performance, such as the attention layer, and  embedding operator. Due to the variation of the valid sequence lengths, it is not easy to implement a general kernel that would perform the best for all cases.

\subsubsection{Load Balance in Distributed Training}
\label{load_balance_issue}

\begin{figure}[!htb]
\centering
\includegraphics[width=0.9\columnwidth]{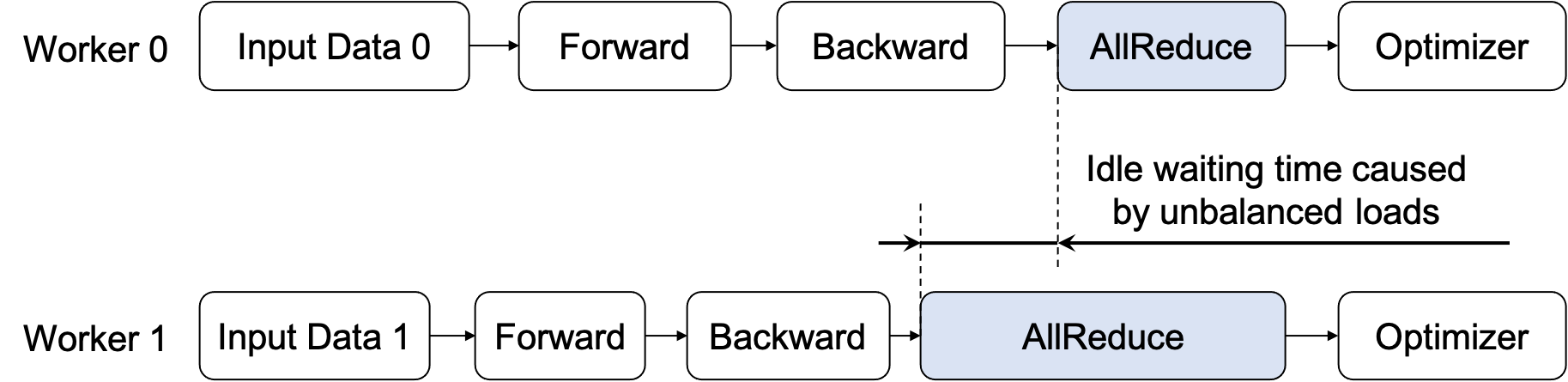}
\caption{The unbalanced loads of the unpadded BERT model.}
\label{unbalanced_allreduce}
\end{figure}

The all-reduce operator is required in the backward part when training the BERT model using the data parallelism distributed strategy. The all-reduce operator is a collective communication primitive, and would not start to launch until all of the workers reach the same time point. As a result, it would cause unnecessary time cost if the workload of each worker is unbalanced, i.e., short board effect, as shown in Figure \ref{unbalanced_allreduce}. The unbalanced workload would be extremely critical when the BERT model takes the variable-length input data. If the padding tokens of the input data are removed, the actual token number to be processed for each worker would be quite different. The different input token numbers to be processed contribute to the different computational time of the forward and backward parts of the model. The workers with less input token number would have to wait for other workers to reach the starting time point of the all-reduce operator. This workload imbalance would harm the speedup ratio in the distributed training process of BERT.

\section{Implementation and Optimization}
This section illustrates two effective methods for inputs of variable length. Besides, a series of systematical optimization methods are also adopted to improve the overall training performance further.

\subsection{Supporting and Optimizing for Variable-length Workloads}
\label{AA}
To achieve high-performance computation for inputs with variable length, we support a broader scope of unpad computing, and also carry out more dedicated optimizations for the most time-consuming multi-head attention layer.

\begin{figure}[!htb]
\centering
\includegraphics[width=0.98\columnwidth]{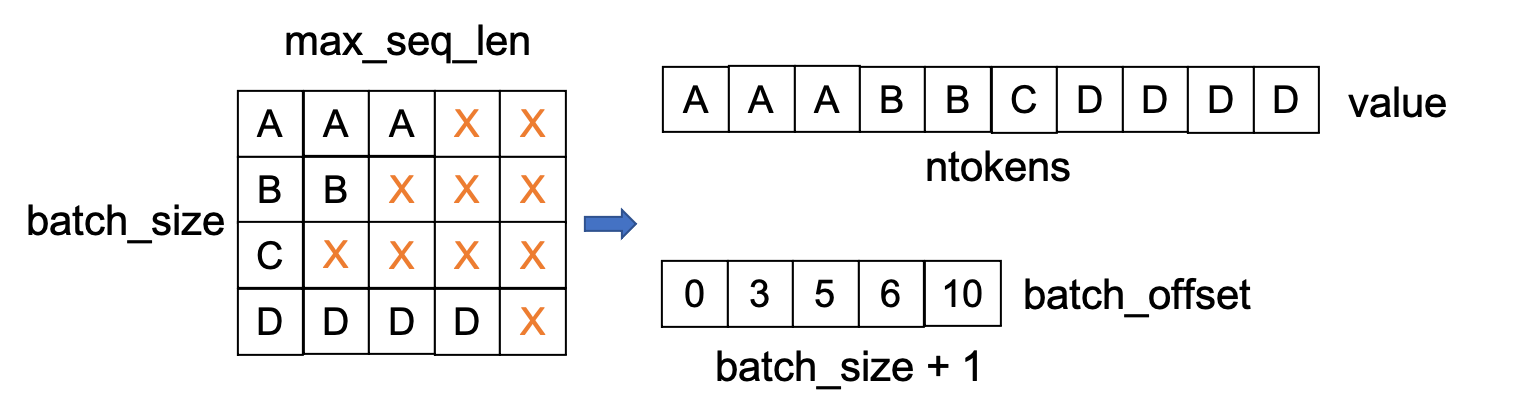}
\caption{The storage for unpad computing. }
\label{fig-storage}
\end{figure}

\subsubsection{Supporting Unpad Computing}
For pad computing, every sequence is padded to a fixed length. As shown in the left of Figure \ref{fig-storage}, the inputs are acting as a two-dimension tensor. To remove the redundant storage space, we merge the \textit{batch\_size} and \textit{max\_seq\_len} dimensions and only store the valid tokens. Besides, a prefix sum array, i.e., \textit{batch\_offset}, is supplied to record the token number of each sequence. 

\begin{figure}[!htb]
\centering
\includegraphics[width=0.62\columnwidth]{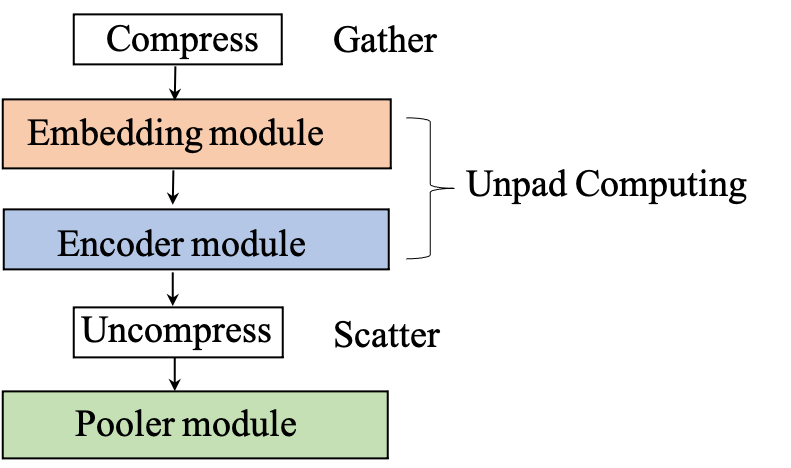}
\caption{Supporting larger scope of unpad computing. }
\label{fig-scope}
\end{figure}

Then we analyze the possibilities to execute unpad computing for the modules in the BERT model. 
For the BERT embedding module, it is natural to execute unpad computing, because the computation of all the involved layers, such as \textit{Embedding}, \textit{LayerNorm}, \textit{Add}, are not affected by merging the \textit{batch\_size} and \textit{max\_seq\_len} dimensions. 
For the BERT encoder module, the dimension of \textit{batch\_size} and \textit{max\_seq\_len} may not be adjacent to each other during the computation process. Thanks to the newly developed FMHA (Fused Multi-Head Attention) kernel \cite{nvidia-apex, flashattention}, it is feasible to execute unpad computing for the encoder module.
For the BERT pooler module, it is unnatural to execute computation after dimension merging. Besides, the pooler module only accounts for a very small amount of time. Therefore, we do not take it into consideration for unpad computing in this paper.

Based on the analysis above, we can execute unpad computing for both BERT embedding and encoder modules, which are the main parts of the BERT model. 
As shown in Figure \ref{fig-scope}, before entering the BERT embedding module, input data are compressed to the unpad format shown in Figure \ref{fig-storage}, which can be obtained by the \textit{gather} operator. 
Before entering the BERT pooler module, the output data from the encoders are uncompressed to the padding format by using a \textit{scatter} operator to avoiding affecting the subsequent calculations. Benefited from the support of unpad computing, the model performance is improved largely by about 2.3x as shown in the experiments.

\subsubsection{Optimizing Unpad Attention Computing}
\begin{figure}[!htb]
\centering
\includegraphics[width=0.98\columnwidth]{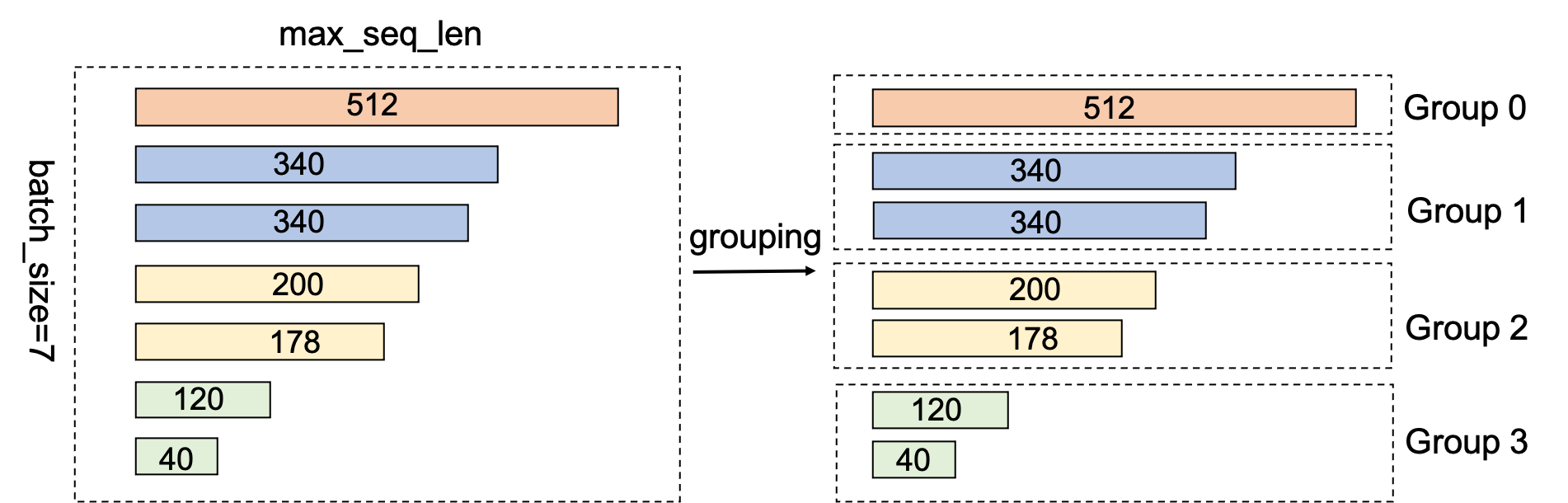}
\caption{An example of grouping method for variable-length sequences. }
\label{fig-group}
\end{figure}

As the most time-consuming part in BERT model, it is vital to improve the performance of the multi-head attention layer. Recently, NVIDIA Apex \cite{nvidia-apex} proposed a highly fused multi-head at-tention kernel, namely, FMHA. It makes the unpad com- puting possible. However, the  existing  FMHA  optimization method is not flexible and effective enough to deal with various sequences with different lengths. As shown on the left of the Figure \ref{fig-group},  no matter what the sequence length distribution is, FMHA assigns a kernel according to the maximum sequence length within a batch, limiting the performance for sequences with small lengths.

To tackle these problems, we propose a grouping and multi-kernel execution method to speed up attention computing. The input sequences within a batch are grouped according to the length based on a criterion. Sequences inside the same group are assigned to a corresponding kernel according to the maximum sequence length within the group.  An example of the grouping method can be seen on the right of the Figure \ref{fig-group}. The six sequences are classified into four groups with the criteria that the sequence lengths are among (0, 128], (128, 256], (256, 384] and (384, 512], respectively. It is noted that the grouping criteria are set according to the underlying kernel implementation and can be adjusted if the FMHA kernel is updated.

\begin{figure}[!htb]
\centering
\includegraphics[width=0.98\columnwidth]{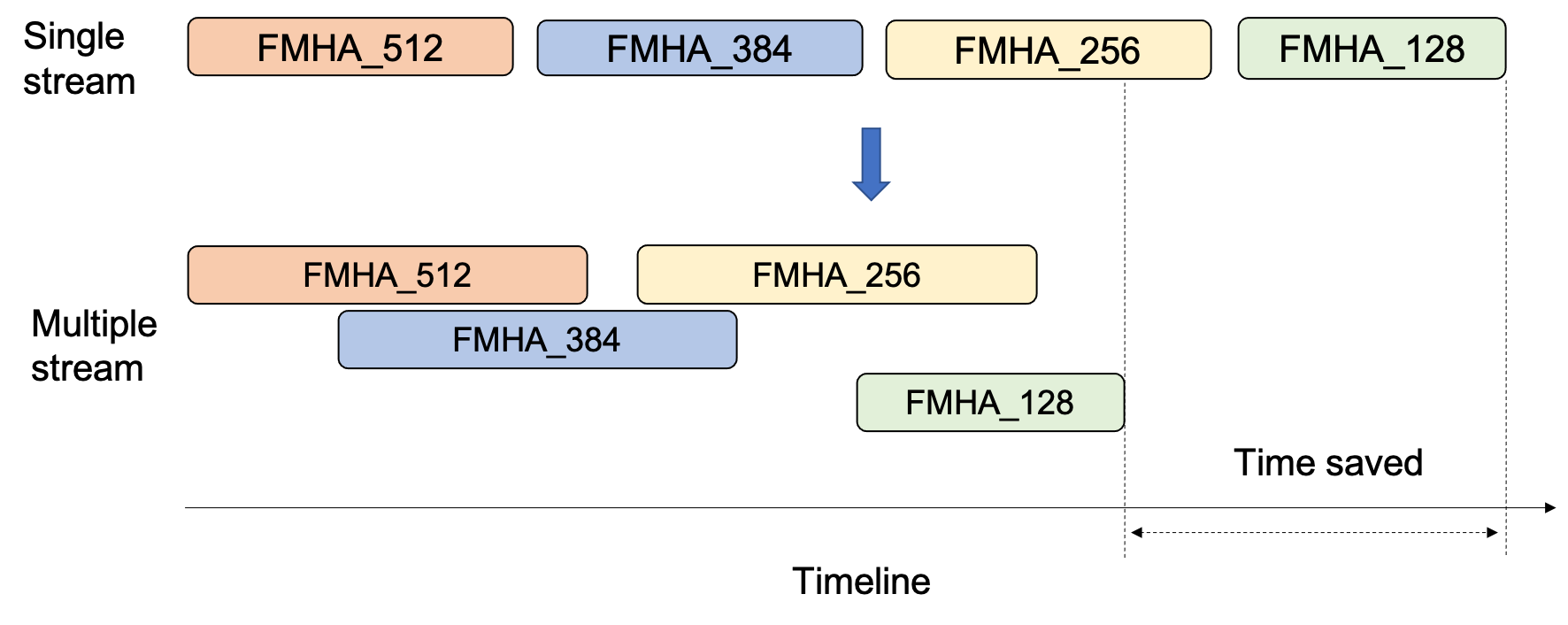}
\caption{Concurrent execution of kernels for different groups. }
\label{fig-fmha-overlap}
\end{figure}
\begin{figure}[!htb]
\centering
\includegraphics[width=0.98\columnwidth]{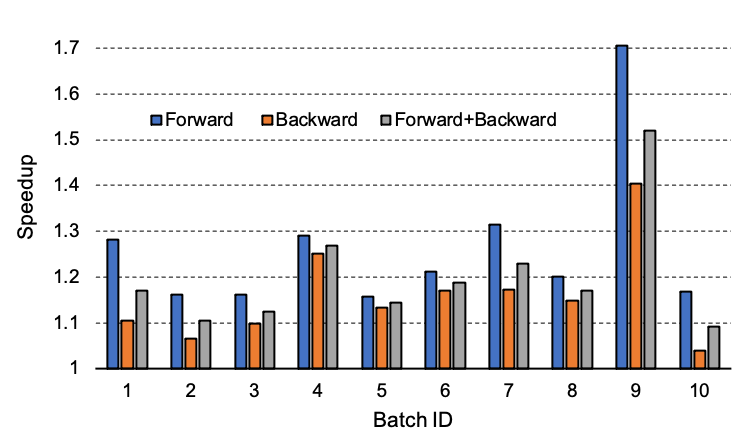}
\caption{Speedup of the grouping and multi-kernel optimization compared with the NVIDIA Apex FMHA. The performance data is from 10 training steps during the training process.}
\label{fig-fmha-perf}
\end{figure}

After grouping, we launch a separate kernel for each group. The launched kernel in each group is determined by the maximum sequence length within the group. Because there are no data dependencies among these kernels, we use multi-stream technology to further improve the GPU resource utilization and performance. As shown in Figure \ref{fig-fmha-overlap},  multiple FMHA kernels are concurrently executed on a GPU by enabling multi-stream optimization, which consumes less time compared with the single stream execution method. Besides, we need to use CUDA events to ensure the execution order of FMHA kernels, and their preceding and subsequent operators to keep the accuracy. All FMHA kernels must start execution after the preceding operator finishes computation, and the subsequent operator must start execution after all of the FMHA kernels finish.
Figure \ref{fig-fmha-perf} shows the speedup of our unpad attention optimization compared with NVIDIA Apex FMHA. Benefited from the grouping and multi-kernel execution, the attention module can bring about 15\%-70\%, 3\%-40\%, and 20\%-52\% performance improvement for forward, backward and overall computations, respectively.

\subsection{Load Balance Optimization in Distributed Training}

As Section \ref{load_balance_issue} addressed, the workload imbalance comes from the different input token numbers to be processed among each worker. The key solution is to make the input token number of each worker nearly the same.

\subsubsection{Padding exchange across the workers} 
\begin{figure}[!htb]
\centering
\includegraphics[width=0.8\columnwidth]{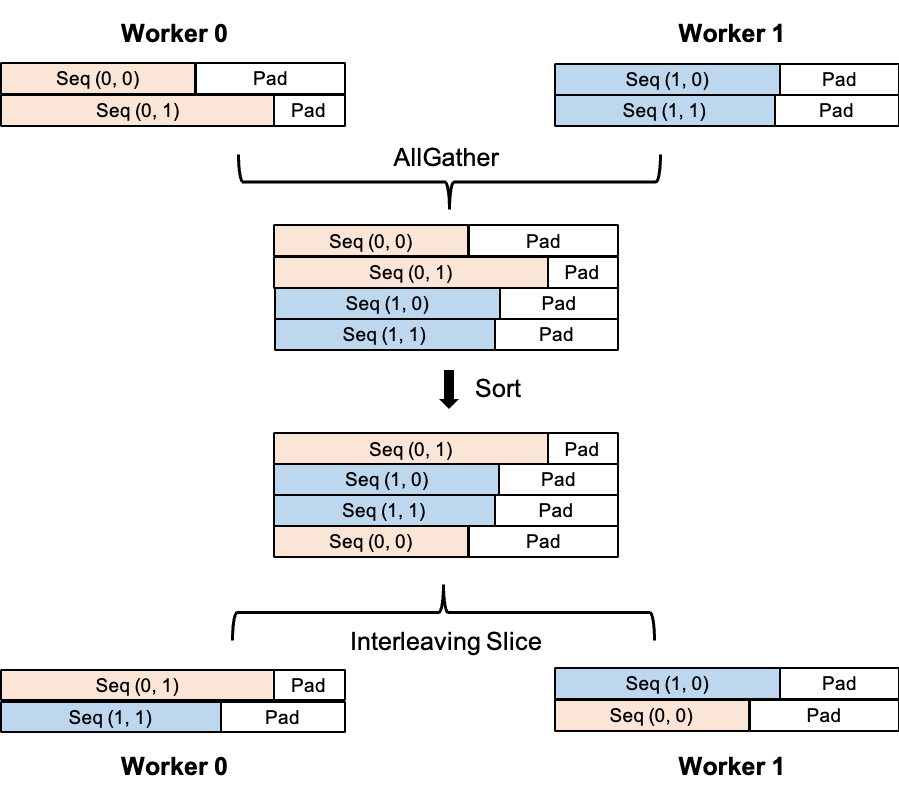}
\caption{The padding exchange process.}
\label{fig-exchange-padding}
\end{figure}

NVIDIA proposed a load balance method named padding exchange in MLPerf Training v1.0\cite{mlperf_benchmark}. The process of the padding exchange method is shown in Figure \ref{fig-exchange-padding}. The detailed steps are as follows:

\begin{itemize}
    \item Perform all-gather operator to concatenate each input tensor across all workers. The BERT model usually contains five input tensors, \textit{input\_ids}, \textit{input\_mask}, \textit{segment\_ids}, \textit{masked\_lm\_labels} and \textit{next\_sentence\_labels}. All of these input tensors contain the padding tokens. Originally, the shape of the input tensors \textit{input\_ids}, \textit{input\_mask}, \textit{segment\_ids} and \textit{masked\_lm\_labels} in each worker is [\textit{batch\_size}, \textit{max\_seq\_len}], while the shape of the input tensor \textit{next\_sentence\_labels} is [\textit{batch\_size}]. Then, input tensors are concatenated  across all workers via the all-gather operator. After the concatenation, each worker would obtain all of the input data in all  of the workers. The shape of \textit{input\_ids}, \textit{input\_mask}, \textit{segment\_ids} and \textit{masked\_lm\_labels} in each worker becomes [\textit{num\_devices} * \textit{batch\_size}, \textit{max\_seq\_len}], while the shape of \textit{next\_sentence\_labels} becomes [\textit{num\_devices} * \textit{batch\_size}]. The reason why the original input tensors would contain padding tokens is that the all-gather operator requires the same data length on each worker. All the workers run the same code and obtain the same results in this step.
    
    \item Sort the all-gathered input sequences along the 0-th dimension according to the valid token number they contain. The valid input token number can be obtained from the input tensor \textit{input\_mask}. All of the workers run the same code and obtain the same results in this step as well.
    
    \item Interleave slicing to obtain the actual input tensor on each worker. The \textit{i}-th worker would retrieve the input sequences from the sorted results using the indices \textit{i}, \textit{i + num\_workers}, \textit{i + 2 * num\_workers}, ... Since the input sequences have been sorted according to the valid input token number, each worker would obtain the different input data with nearly the same input token number after this step.
\end{itemize}

The padding exchange method proposed by NVIDIA has the following disadvantages:

\begin{itemize}
    \item All of the steps are performed using the GPU operators. During the BERT training process, the GPU workload is usually extremely high, for example the utilization rate would be larger than 98\%. These GPU operators in the padding exchange method would occupy the computational resources of GPU, which results in a performance drop.
    \item The padding exchange process and the model training process execute serially. The padding exchange method is performed right at the beginning of each mini-batch training.
\end{itemize}

\begin{figure}[!htb]
\centering
\includegraphics[width=0.80\columnwidth]{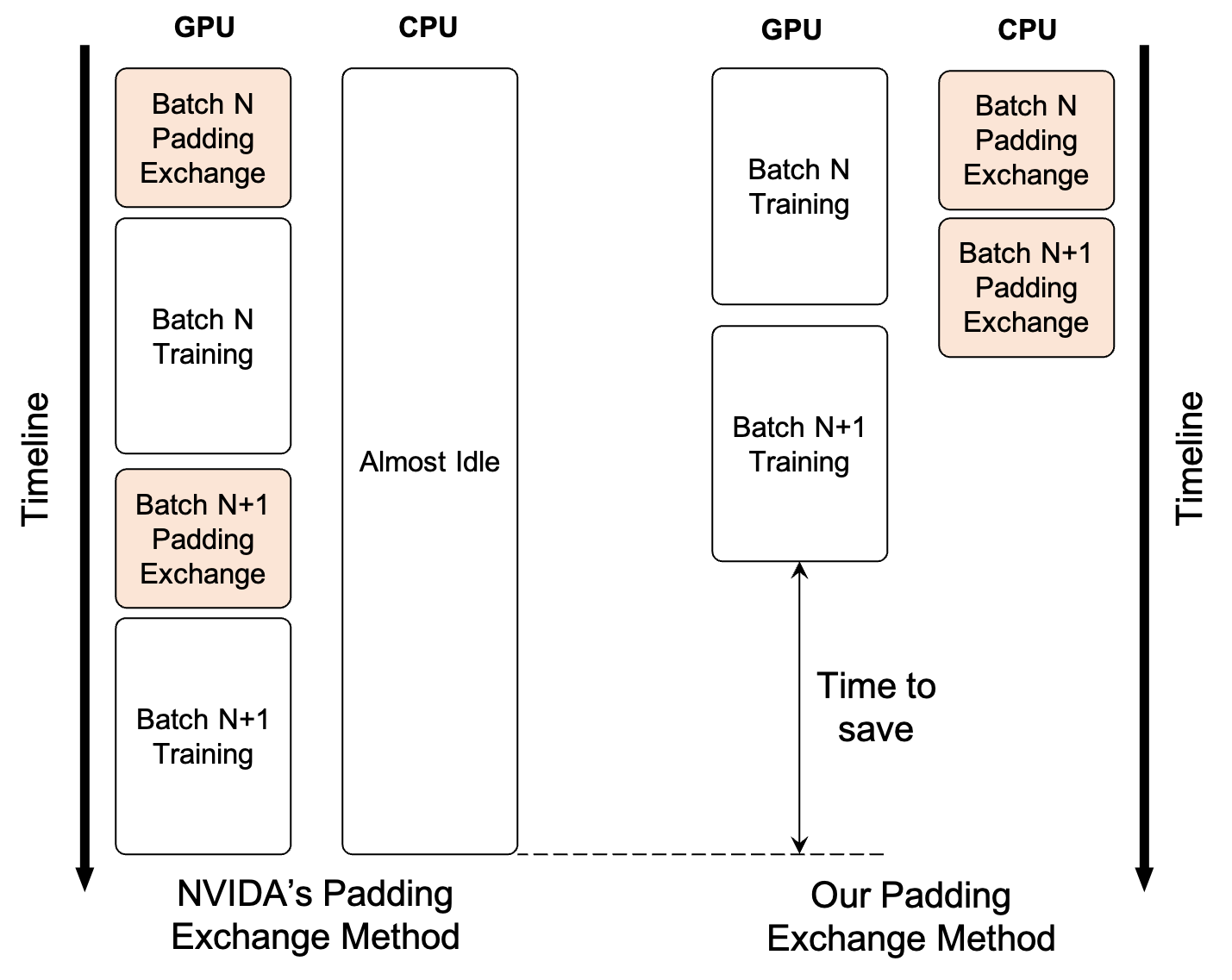}
\caption{Overlapping of padding exchange and GPU computation.}
\label{fig-cpu-exchange-padding}
\end{figure}

Inspired by the NVIDIA's implementation, we proposed our optimized padding exchange method in this paper, as shown in Figure \ref{fig-cpu-exchange-padding}. The highlights of the optimized method in this paper are as follows:

\begin{itemize}
    \item Use the CPU to perform the padding exchange process instead of GPU. During the BERT training process, the GPU is always busy, but the CPU almost idles all the time. All of the GPU operators above can be moved to the CPU side and the NCCL all-gather operator is replaced with the MPI alternatives.
    \item Pre-perform the padding exchange method to prepare the following mini-batch data when the network is training on GPU. Since the CPU and GPU are two individual devices, the padding exchange process on the CPU and the training process on GPU can overlap. The time cost of the padding exchange is usually much smaller than the time cost of training one mini-batch. Therefore, the next mini-batch data is always ready before the next mini-batch training starts.
    \item Overlap the GPU training process with the host-to-device memory copy of the next mini-batch data. The optimized padding exchange method in this paper would generate the CPU tensor instead of the GPU tensor like NVIDIA’s method. This paper uses the multiple CUDA streams methodology to perform host-to-device memory copy asynchronously with the GPU training process.
\end{itemize}

\subsubsection{Overlapping of the Padding Removal Process and the GPU Training}

Due to the requirement of the all-gather operator, the input data before and after the padding exchange process have to contain the padding tokens. The padding tokens should be removed before running the following network layers.  Solutions include: 1. obtain the index tensor \textit{nonzero\_indices} where the input tensor \textit{input\_mask} is non-zero, 2.  run the \textit{gather} operator to remove the padding of the input tensors.

In this paper, we calculate the \textit{nonzero\_indices} during the CPU padding exchange process due to: 

\begin{itemize}
    \item If we calculate the \textit{nonzero\_indices} tensor after the padding exchange process, the \textit{input\_mask} is a GPU tensor. Since we cannot know the final shape of the \textit{nonzero\_indices} tensor on the host side, we must perform host-device synchronization to get the shape of the \textit{nonzero\_indices} tensor first. This host-device synchronization can be removed if we calculate the \textit{nonzero\_indices} during the CPU padding exchange process.
    \item The padding exchange process can fully overlap with the GPU training process, so that the calculation and H2D copy of the \textit{nonzero\_indices} tensor can be done asynchronously when the BERT model is training in the meanwhile.
\end{itemize}

Generally, all the operators that are only related to the input tensors can run during the CPU padding exchange process to overlap more. It includes the \textit{gather} operator to remove the padding tokens mentioned above, and the generation of the \textit{batch\_offset} tensor in Figure \ref{fig-storage}.

\subsection{Other Computational Optimizations}

\subsubsection{Kernel Fusion}
Kernel fusion is a common technique used in deep learning to increase the computation efficiency by reducing the number of global memory accesses, increasing data locality, and reducing kernel launch overhead. In this work, we use kernel fusion for more broader computation patterns, such as Linear, Linear\_GeLU\_Linear, Dropout\_Add\_LayerNorm, and Grad of Residual block. The detailed kernel fusion is as followings.

\begin{itemize}
\item Linear Fusion: Linear is one of the most frequently  called operators in the BERT model. It is usually composed of a GEMM and a bias-adding computation. The bias-adding computation usually needs a broadcast of the bias tensor in the BERT model, whose backward corresponds to the reduction computation pattern. Thanks to the support of the cuBLASLt \cite{cublaslt} library, it is possible to fuse the forward kernels (GEMM and bias-adding) and their corresponding backward kernels (GEMM and reduction).

\item Linear\_GeLU\_Linear Fusion: Apart from the Linear fusion, cuBLASLt also supports this fusion with an additional GeLU or ReLU computation, for both forward and backward. This feature can be used to accelerate the Linear\_GeLU\_Linear pattern in BERT Encoder. By using this fusion, the number of kernels is reduced from 12 to 6.

\item Dropout\_Add\_LayerNorm Fusion: There are also some \textit{Dropout}, \textit{Add}, and \textit{LayerNorm} patterns in the BERT encoder. For \textit{Dropout} and \textit{Add} operators, they all belong to the elementwise pattern, which can be naturally fused to the subsequent \textit{LayerNorm} kernels. From the parallel optimization, we learn the parallel task distribution idea from \cite{nvidia-apex}, and fuse these three operators into one kernel for the forward part and two for the backward part. 

\item Fusion of Residual Grad: Like the ResNet model \cite{resnet}, there are two residual blocks in the BERT encoder. In the BERT model, the residual tensor is not only the input of the Linear layer, but also the input of the other subsequent layer, resulting to an additional adding computation to get the gradient of residual. Aiming to remove this explicit adding computation kernel, we make use of the \textit{beta} parameter of BLAS GEMM API \cite{blas-gemm} to fuse the gradient addition kernel with the GEMM kernel in the backward part of Linear.
\end{itemize}

Table \ref{table:kernel-fusion} summarizes kernel number changes by using the kernel fusion technique. It is evident that the kernel numbers for the above patterns have been reduced largely, with roughly 1.6x to 2.6x.

\tabcolsep 1.4pt
\begin{table}[!h]
\scriptsize
\footnotesize
\addtolength{\tabcolsep}{2.0pt}
\renewcommand{\arraystretch}{1.25}
\caption{The kernel numbers changes by using kernel fusion technique.}
\label{table:kernel-fusion}
\centering
\begin{tabular}{lccc}
 \hline
 Patterns   & Forward  & Backward & Forward\&Backward  \\
 \hline
  Linear & 2 $\rightarrow$ 1 & 3 $\rightarrow$ 2 & 5 $\rightarrow$ 3   \\
  \hline
  Linear\_GeLU\_Linear & 5 $\rightarrow$ 2 & 7 $\rightarrow$ 4 & 12 $\rightarrow$ 6  \\
  \hline
Dropout\_Add\_LayerNorm & 3 $\rightarrow$ 1 & 5 $\rightarrow$ 2 & 8 $\rightarrow$ 3     \\
  \hline
  Residual Grad & - & 2 $\rightarrow$ 1 & 2 $\rightarrow$ 1 \\
  \hline
  \end{tabular}
\end{table}

\subsubsection{LAMB Optimization}

Our LAMB optimizer implementation starts from the \textit{DistributedFusedLAMB} API in the NVIDIA Apex\cite{nvidia-apex} library. All the FP16/FP32 parameters, the FP32 master parameters, the FP16/FP32 gradients of the parameters, the trust ratio tensor, and the FP32 momentum of the LAMB optimizer are flattened and copied into tensors with contiguous memory space respectively.

The LAMB optimizer needs to calculate the following tensors:

\begin{itemize}\label{lamb-3-case-tensor}
    \item Case 1: the squared sum of the L2-Norm of each gradient. The L2-Norm of the gradients is used to perform global norm clipping. The result is a tensor with shape [1].
    \item Case 2: the L2-Norm of each parameter. The result is a tensor with shape \textit{num\_parameters}, where \textit{num\_parameters} is the parameter number of the model.
    \item Case 3: the L2-Norm of each of the trust ratio tensors. The result is a tensor with shape [\textit{num\_parameters}].
\end{itemize}

In the NVIDIA Apex library, all three tensors above are calculated by the \textit{multi\_tensor\_apply} function. Given the size of a chunk, i.e., \textit{chunk\_size}, each of the input tensors is divided into several chunks, where the chunk number is \textit{ceil(tensor.numel() / chunk\_size)}. The information of each tensor’s chunk is recorded in a data structure named \textit{TensorListMetadata} \ref{code-tensor-list-meta-data} (pseudo code) sequentially. The maximum tensor number and block number inside the \textit{TensorListMetadata} are limited by the template argument \textit{MaxTensorNum} and \textit{MaxChunkNum} respectively. When the tensor number or the block number exceeds its limit, a new CUDA kernel would be launched with the \textit{TensorListMetadata} object as its argument. Each thread in the CUDA kernel would process the data from the same chunk, and the block number of the CUDA kernel is the same with the chunk number. Several CUDA kernels may be launched in order to process all the chunks of the tensors.

\begin{lstlisting}[language=c++, label=code-tensor-list-meta-data]
template <int MaxTensorNum, int MaxChunkNum> 
struct TensorListMetadata
{
  void* addresses[MaxTensorNum];
  int sizes[MaxTensorNum];
  unsigned char block_to_tensor[MaxChunkNum];
  int block_to_chunk[MaxChunkNum];
  int start_tensor_this_launch;
};
\end{lstlisting}

Since the \textit{TensorListMetadata} object is the argument of the CUDA kernel, the \textit{MaxTensorNum} and \textit{MaxChunkNum} should not be too large because the memory size of the CUDA kernel arguments is limited by CUDA (for example, 4KB on A100 GPU). Therefore, the maximum chunk number to be processed in a CUDA kernel is limited by the \textit{sizeof(TensorListMetadata)} . It is important to increase the \textit{MaxTensorNum} and \textit{MaxChunkNum}, so that we can launch fewer CUDA kernels to achieve more parallelism. We found that due to the large \textit{sizeof(TensorListMetadata)}, the \textit{DistributedFusedLAMB} optimizer in Apex would launch 1 to 5 CUDA kernels to calculate each of the 3 tensors mentioned above \ref{lamb-3-case-tensor} respectively.

In our paper, we improve the performance of the \textit{multi\_tensor\_apply} method in the \textit{DistributedFusedLAMB} optimizer. It should be noticed that, all of the gradients, parameters and trust ratio tensors are flattened and copied to a tensor with contiguous memory space. Therefore, we can only record the data pointer of the contiguous memory tensor inside \textit{TensorListMetadata} instead of an array of each tensor \textit{TensorListMetadata::address}. In this way, the \textit{sizeof(TensorListMetadata)} would be smaller, and we can increase the \textit{MaxTensorNum} and \textit{MaxChunkNum}, so that the calculation of \ref{lamb-3-case-tensor} can be completed in one CUDA kernel respectively. Particularly, the tensor in case 1 \ref{lamb-3-case-tensor} can be calculated using the \textit{cub::DeviceReduce::Reduce} method in NVIDIA cub library instead of the \textit{multi\_tensor\_apply} method above for better performance. Table \ref{table:lamb_optimization} presents the performance of the \textit{DistributedFusedLAMB} optimizer in the BERT-Large model. Our optimized version is about 22\% faster than the NVIDIA Apex.

\tabcolsep 1.4pt
\begin{table}[!h]
\scriptsize
\footnotesize
\addtolength{\tabcolsep}{2.0pt}
\renewcommand{\arraystretch}{1.25}
\caption{DistributedFusedLAMB Performance Compared with the NVIDIA Apex}
\label{table:lamb_optimization}
\centering
\begin{tabular}{lcc}
  \hline
  Implementation   & Time cost per step (ms) \\
  \hline
  NVIDIA Apex Library    &   10.68 \\
  \hline
  Ours            &   8.30 \\
  \hline
  \end{tabular}
\end{table}
 
\subsubsection{Embedding Operator Optimization}
The embedding module in the BERT model is used to learn the vector representation of the input tokens. In the forward part, the index tensor is used to look up the weight tensor and generate the output tensor. In the backward part, the output's gradient from the same index should be accumulated to get the weight's gradient. However, if different threads deal with the workloads from the same index, there will be write conflicts.  With the development of hardware, the performance of \textit{atomicAdd} instruction is becoming more and more effective. Therefore, TensorFlow \cite{tensorflow} and PaddlePaddle \cite{paddle} use \textit{atomicAdd} instruction to deal with the write conflicts directly. As for PyTorch \cite{pytorch}, it depends on a sort-based method to eliminate the conflicts, which is composed of multiple CUDA kernels. 

However, the existing methods still have some disadvantages:
\begin{itemize}
    \item Workload distribution problem: For the TensorFlow embedding layer, the task distribution is not effective enough, where the workload of each thread is too heavy. As for the PaddlePaddle embedding layer, the number of spawn blocks is too small, even less than the number of the SMs in the GPU. All of these problems can lead to low GPU utilization and bad performance.
    \item \textit{AtomicAdd} performance problem: Compared with the sort-based method used in PyTorch embedding, the \textit{atomicAdd} method has the potential to be faster. However, it is found the performance of \textit{atomicAdd} instruction for FP16 type is abnormal, even making performance worse than the sort-based method, which may be caused by the poor hardware support.
\end{itemize}

\begin{figure}[!htb]
\centering
\includegraphics[width=0.98\columnwidth]{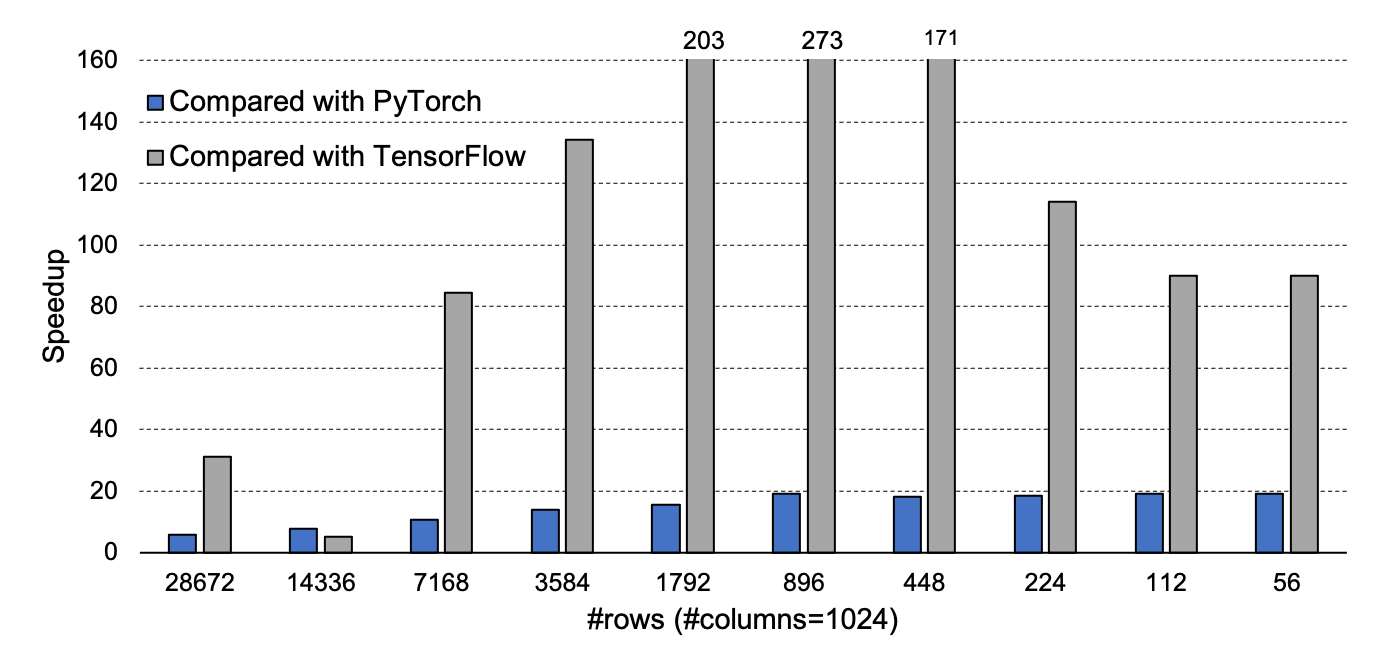}
\caption{Speedup of embedding backward computation for FP16 data type on Tesla A100.}
\label{fig-embed-perf}
\end{figure}

To solve the workload distribution problem, we adjust the thread configurations properly, by increasing the number of blocks and making the workloads of one thread lighter.
For \textit{atomicAdd} performance problem, it is lucky to find that the performance of \textit{atomicAdd} for half2 data type is normal and effective. Therefore, it is natural to convert FP16 to half2 type to improve the performance. Specifically, two adjacent FP16 elements along the embedding weight column are packed into a half2 element, and then \textit{atomicAdd(half2* )} is called to execute the accumulation. Our optimization is implemented on the basis of PaddlePaddle. Experiments in Figure \ref{fig-embed-perf} show that the performance of embedding backward achieve 5.7x-19.2x and 5.1x-273.1x speedups over the PyTorch and TensorFlow implementations, respectively.

\subsubsection{Reduce CPU and GPU Synchronization}
During training and evaluation, the interactions between CPU and GPU introduce some synchronization overheads. For example, learning rates are computed on the CPU, but consumed on GPU. Thus, an additional H2D (Host to Device) memory copy is needed to transfer learning rate values. Besides, multiple D2H (Device to Host) memory copies are executed  to fetch some useful information, e.g., loss and accuracy, in every iteration. To reduce synchronizations, we first move the computation of the learning rate to GPU and eliminate the H2D memory copies. In addition, we control the frequency of the information collection as lower as possible. For example, if it is not necessary to acquire information every iteration step, we can invoke it only every \textit{n} steps, where \textit{n} is a positive integer. By this technique, the D2H memory copies can be reduced vastly.

\section{Experiments}

We tested the performance of our optimization methods on the MLPerf BERT-Large model. We adopted the deep learning framework PaddlePaddle\cite{paddle} to run our optimization. The tests were running on 8x NVIDIA A100 of 400W. The global batch size of all tests was 448 (56 batch size per GPU * 8 GPUs) and the gradient accumulation step was 1. All the tests were running under the O2-level mixed-precision method with the LAMB optimizer.

\subsection{Performance of Different Optimizations}
\begin{figure}[!htb]
\centering
\includegraphics[width=0.98\columnwidth]{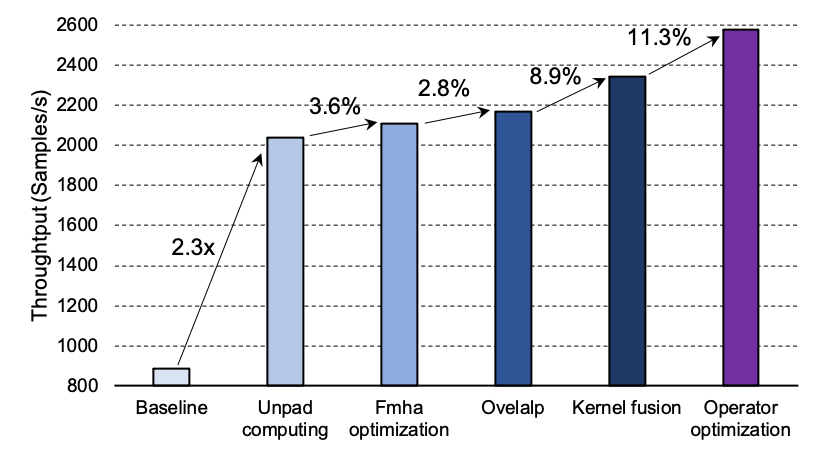}\caption{Performance for different optimization techniques.}
\label{fig-optimization-break}
\end{figure}
To demonstrate the proposed efficiency of the optimizations, we first conducted experiments to test the performance improvements breakdown for different optimizations. The baseline was the implementation with the typical padding methods. By supporting unpad computing for the embedding and encoder modules, we achieved around 2.3x speedup over the baseline. On top of that, the grouping and multiple-stream optimizations for unpad 
FMHA operator brought about 3.6\% performance improvement. Besides, the overlapping of padding exchange and removal process with computations on GPU could get about a 2.8\% improvement. By kernel fusions on a series of computation patterns, we also obtained an 8.9\% improvement in performance. Besides, operator optimizations for LAMB and embedding brought about 11.3\% improvements.

\subsection{Speedup Ratios of Distributed Training}
\begin{figure}[!htb]
\centering
\includegraphics[width=0.92\columnwidth]{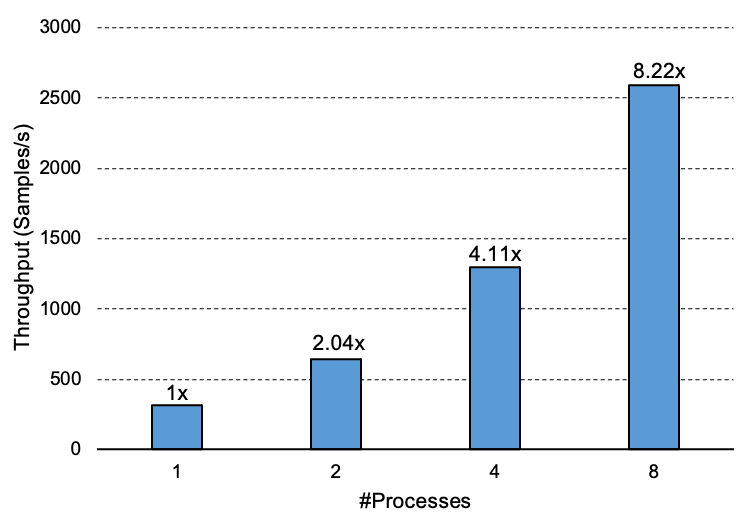}
\caption{The speedup ratio of the BERT model in distributed training.}
\label{fig-scalability}
\end{figure}
To test the effectiveness of the distributed strategy and optimizations, we also made some experiments to show the speedup ratios in distributed training. We fixed the inputs and adjusted the number of the processes from 1 to 8 to observe the throughput changes. As shown in Figure \ref{fig-scalability}, with the increase of the processes, the throughput would improve stably, even with super-linear acceleration. This is mainly benefited from the proposed distributed training optimization in the LAMB operator.

\subsection{End-to-End Performance}

\subsubsection{Throughput Comparison} 
\tabcolsep 1.4pt
\begin{table}[!h]
\scriptsize
\footnotesize
\addtolength{\tabcolsep}{2.0pt}
\renewcommand{\arraystretch}{1.25}
\caption{Throughput comparison of the mainstream BERT implementation.}
\label{table:bert_throughput}
\centering
\begin{tabular}{lcc}
 \hline
 BERT Model   & Throughput (Samples/s) \\
  \hline
  NVIDIA MLPerf BERT\cite{mlperf-2.0}    &   2415 \\
  \hline
  HazyResearch MLPerf BERT\cite{mlperf-2.0}    &   2424 \\
  \hline
  DeepSpeed BERT\cite{deepspeed-bert}   &   876 \\
  \hline
  Megatron-LM BERT\cite{megatron-lm}  &   819 \\
  \hline
  Ours                        &   2578 \\
  \hline
  
  \end{tabular}
\end{table}

We also made comparisons with many mainstream BERT implementations, such as the DeepSpeed, Megatron-LM and two representative MLPerf Training v2.0 results from NVIDIA and HazyReasearch. As shown in Table \ref{table:bert_throughput}, our work can achieve the fastest throughput with roughly 2578 Samples/s. Compared with DeepSpeed and Megatron-LM, which carry out padding computing, we can achieve above 2.9x speedup. Most of benefits are from the removal of the redundant computing. For NVIDIA and HazyResearch MLPerf results, they also used unpad computing on BERT encoder module. Compared with them, we can get about 6\% improvements. This is mainly benefited from more broader and deeper optimization on top of the unpad computing, such as the unpad FMHA optimization, overlaps, operator optimization and so on.

\subsubsection{MLPerf Training v2.0 Performance}
We submitted our work to the international {MLPerf training v2.0 benchmark \cite{mlperf-2.0}}, which provides a unified platform to compare performance of various models from different fields. The BERT-Large model used for pre-training is selected as one of the reference models in the MLPerf model suites. The machine we tested was with 8x NVIDIA A100 of 400W. The metrics for the MLPerf BERT model is the time to train that converges to 72\% MLM (Masked Language Modeling) accuracy on the evaluation data set. The final results are listed in Table \ref{table:mlperf-results}. As shown in the table, our work can converge with the minimum time, i.e., 16.598 minutes. Compared with other submitted works, we can achieve 1.05x-1.11x speedups, demonstrating the effectiveness of all the optimizations.

\tabcolsep 1.4pt
\begin{table}[!h]
\scriptsize
\footnotesize
\addtolength{\tabcolsep}{2.0pt}
\renewcommand{\arraystretch}{1.25}
\caption{Results of MLPerf Training v2.0 for BERT on 8x NVIDIA A100 400W GPUs.}
\label{table:mlperf-results}
\centering
\begin{tabular}{lcc}
 \hline
  Submitter   & End to End Time (minutes) \\
 \hline
  GIGABYTE       & 18.489      \\
  \hline
   NVIDIA         & 18.442     \\
  \hline
  H3C            & 17.623      \\
  \hline
   HazyResearch   & 17.402     \\
  \hline
  Ours            & 16.598      \\
  \hline
  \end{tabular}
\end{table}

\section{Conclusion}
\label{sec-conclusion}
In this paper, we proposed our novel performance optimization of the unpadded BERT model with varible-length inputs. The redundant computations from the padding tokens are removed to improve the training performance. The architecture of the model with variable-length inputs is proposed and we make some efficient optimizations to improve the distributed training performance, including grouped multi-stream FMHA, data balance overlapped with the training process, and other CUDA kernel optimizations. These optimizations achieve the state-of-the-art performance of the BERT-Large model. We would be dedicated to apply our optimizations on other Transformer-based models with variable-length inputs in the future.

\bibliographystyle{IEEEtran}
\bibliography{bert_paddle}

\begin{thebibliography}{10}
\providecommand{\url}[1]{#1}
\csname url@samestyle\endcsname
\providecommand{\newblock}{\relax}
\providecommand{\bibinfo}[2]{#2}
\providecommand{\BIBentrySTDinterwordspacing}{\spaceskip=0pt\relax}
\providecommand{\BIBentryALTinterwordstretchfactor}{4}
\providecommand{\BIBentryALTinterwordspacing}{\spaceskip=\fontdimen2\font plus
\BIBentryALTinterwordstretchfactor\fontdimen3\font minus
  \fontdimen4\font\relax}
\providecommand{\BIBforeignlanguage}[2]{{%
\expandafter\ifx\csname l@#1\endcsname\relax
\typeout{** WARNING: IEEEtran.bst: No hyphenation pattern has been}%
\typeout{** loaded for the language `#1'. Using the pattern for}%
\typeout{** the default language instead.}%
\else
\language=\csname l@#1\endcsname
\fi
#2}}
\providecommand{\BIBdecl}{\relax}
\BIBdecl

\bibitem{vaswani2017attention}
A.~Vaswani, N.~Shazeer, N.~Parmar, J.~Uszkoreit, L.~Jones, A.~N. Gomez,
  {\L}.~Kaiser, and I.~Polosukhin, ``Attention is all you need,''
  \emph{Advances in neural information processing systems}, vol.~30, 2017.

\bibitem{2018-bert}
J.~Devlin, M.-W. Chang, K.~Lee, and K.~Toutanova, ``Bert: Pre-training of deep
  bidirectional transformers for language understanding,'' \emph{arXiv preprint
  arXiv:1810.04805}, 2018.
  
\bibitem{sun2019ernie}
\BIBentryALTinterwordspacing
Z.~Zhang, X.~Han, Z.~Liu, X.~Jiang, M.~Sun, and Q.~Liu, ``{ERNIE}: Enhanced
  language representation with informative entities,'' in \emph{Proceedings of
  the 57th Annual Meeting of the Association for Computational
  Linguistics}.\hskip 1em plus 0.5em minus 0.4em\relax Florence, Italy:
  Association for Computational Linguistics, Jul. 2019, pp. 1441--1451.
  [Online]. Available: \url{https://aclanthology.org/P19-1139}
\BIBentrySTDinterwordspacing

\bibitem{liu2019roberta}
Y.~Liu, M.~Ott, N.~Goyal, J.~Du, M.~Joshi, D.~Chen, O.~Levy, M.~Lewis,
  L.~Zettlemoyer, and V.~Stoyanov, ``Roberta: A robustly optimized bert
  pretraining approach,'' \emph{arXiv preprint arXiv:1907.11692}, 2019.

\bibitem{yang2019xlnet}
Z.~Yang, Z.~Dai, Y.~Yang, J.~Carbonell, R.~R. Salakhutdinov, and Q.~V. Le,
  ``Xlnet: Generalized autoregressive pretraining for language understanding,''
  \emph{Advances in neural information processing systems}, vol.~32, 2019.

\bibitem{clark2020electra}
K.~Clark, M.-T. Luong, Q.~V. Le, and C.~D. Manning, ``Electra: Pre-training
  text encoders as discriminators rather than generators,'' \emph{arXiv
  preprint arXiv:2003.10555}, 2020.

\bibitem{zhai2022scaling}
X.~Zhai, A.~Kolesnikov, N.~Houlsby, and L.~Beyer, ``Scaling vision
  transformers,'' in \emph{Proceedings of the IEEE/CVF Conference on Computer
  Vision and Pattern Recognition}, 2022, pp. 12\,104--12\,113.

\bibitem{gulati2020conformer}
A.~Gulati, J.~Qin, C.-C. Chiu, N.~Parmar, Y.~Zhang, J.~Yu, W.~Han, S.~Wang,
  Z.~Zhang, Y.~Wu \emph{et~al.}, ``Conformer: Convolution-augmented transformer
  for speech recognition,'' \emph{arXiv preprint arXiv:2005.08100}, 2020.

\bibitem{brown2020language}
T.~Brown, B.~Mann, N.~Ryder, M.~Subbiah, J.~D. Kaplan, P.~Dhariwal,
  A.~Neelakantan, P.~Shyam, G.~Sastry, A.~Askell \emph{et~al.}, ``Language
  models are few-shot learners,'' \emph{Advances in neural information
  processing systems}, vol.~33, pp. 1877--1901, 2020.

\bibitem{rajbhandari2020zero}
S.~Rajbhandari, J.~Rasley, O.~Ruwase, and Y.~He, ``Zero: Memory optimizations
  toward training trillion parameter models,'' in \emph{SC20: International
  Conference for High Performance Computing, Networking, Storage and
  Analysis}.\hskip 1em plus 0.5em minus 0.4em\relax IEEE, 2020, pp. 1--16.

\bibitem{rasley2020deepspeed}
J.~Rasley, S.~Rajbhandari, O.~Ruwase, and Y.~He, ``Deepspeed: System
  optimizations enable training deep learning models with over 100 billion
  parameters,'' in \emph{Proceedings of the 26th ACM SIGKDD International
  Conference on Knowledge Discovery \& Data Mining}, 2020, pp. 3505--3506.

\bibitem{shoeybi2019megatron}
M.~Shoeybi, M.~Patwary, R.~Puri, P.~LeGresley, J.~Casper, and B.~Catanzaro,
  ``Megatron-lm: Training multi-billion parameter language models using model
  parallelism,'' \emph{arXiv preprint arXiv:1909.08053}, 2019.
  
\bibitem{lightseq-train}
X.~Wang, Y.~Xiong, X.~Qian, Y.~Wei, L.~Li, and M.~Wang, ``Lightseq2:
  Accelerated training for transformer-based models on gpus,'' \emph{arXiv
  preprint arXiv:2110.05722}, 2021.

\bibitem{yan2021fastseq}
Y.~Yan, F.~Hu, J.~Chen, N.~Bhendawade, T.~Ye, Y.~Gong, N.~Duan, D.~Cui, B.~Chi,
  and R.~Zhang, ``{F}ast{S}eq: Make sequence generation faster,'' in
  \emph{Proceedings of the 59th Annual Meeting of the Association for
  Computational Linguistics and the 11th International Joint Conference on
  Natural Language Processing: System Demonstrations}, 2021.

\bibitem{turbotransformers}
J.~Fang, Y.~Yu, C.~Zhao, and J.~Zhou, ``Turbotransformers: an efficient gpu
  serving system for transformer models,'' in \emph{Proceedings of the 26th ACM
  SIGPLAN Symposium on Principles and Practice of Parallel Programming}, 2021,
  pp. 389--402.

\bibitem{mlperf_benchmark}
\BIBentryALTinterwordspacing
P.~Mattson, C.~Cheng, G.~Diamos, C.~Coleman, P.~Micikevicius, D.~Patterson,
  H.~Tang, G.-Y. Wei, P.~Bailis, V.~Bittorf, D.~Brooks, D.~Chen, D.~Dutta,
  U.~Gupta, K.~Hazelwood, A.~Hock, X.~Huang, D.~Kang, D.~Kanter, N.~Kumar,
  J.~Liao, D.~Narayanan, T.~Oguntebi, G.~Pekhimenko, L.~Pentecost,
  V.~Janapa~Reddi, T.~Robie, T.~St~John, C.-J. Wu, L.~Xu, C.~Young, and
  M.~Zaharia, ``Mlperf training benchmark,'' in \emph{Proceedings of Machine
  Learning and Systems}, I.~Dhillon, D.~Papailiopoulos, and V.~Sze, Eds.,
  vol.~2, 2020, pp. 336--349. [Online]. Available:
  \url{https://proceedings.mlsys.org/paper/2020/file/02522a2b2726fb0a03bb19f2d8d9524d-Paper.pdf}
\BIBentrySTDinterwordspacing

\bibitem{mlperf-2.0}
MLCommons, ``{{MLPerf Training v2.0 Results}},'' Website, 2022,
  \url{https://mlcommons.org/en/training-normal-20/}.

\bibitem{nvidia-apex}
NVIDIA, ``Nvidia apex library,'' Website, 2022,
  \url{https://github.com/NVIDIA/apex}.
  
\bibitem{flashattention}
T.~Dao, D.~Y. Fu, S.~Ermon, A.~Rudra, and C.~R{\'e}, ``Flashattention: Fast and
  memory-efficient exact attention with io-awareness,'' \emph{arXiv preprint
  arXiv:2205.14135}, 2022.


\bibitem{cublaslt}
NVIDIA, ``{CuBLASLt API},'' Website, 2022,
  \url{https://docs.nvidia.com/cuda/cublas/index.html#using-the-cublasLt-api}.

\bibitem{resnet}
K.~He, X.~Zhang, S.~Ren, and J.~Sun, ``Deep residual learning for image
  recognition,'' in \emph{Proceedings of the IEEE Conference on Computer Vision
  and Pattern Recognition (CVPR)}, June 2016.

\bibitem{blas-gemm}
NVIDIA, ``{CuBLAS API},'' Website, 2022,
  \url{https://docs.nvidia.com/cuda/cublas/index.html}.

\bibitem{tensorflow}
T.~community, ``Tensorflow: An open source machine learning framework for
  everyone,'' Website, 2022, \url{https://github.com/tensorflow/tensorflow}.
  
\bibitem{paddle}
\BIBentryALTinterwordspacing
T.~W. H.~W. Yanjun~Ma, Dianhai~Yu, ``Paddlepaddle: An open-source deep learning
  platform from industrial practice,'' \emph{Frontiers of Data and Domputing},
  vol.~1, no.~1, p. 105, 2019. [Online]. Available:
  \url{http://www.jfdc.cnic.cn/EN/abstract/article\_2.shtml}
\BIBentrySTDinterwordspacing

\bibitem{pytorch}
\BIBentryALTinterwordspacing
A.~Paszke, S.~Gross, F.~Massa, A.~Lerer, J.~Bradbury, G.~Chanan, T.~Killeen,
  Z.~Lin, N.~Gimelshein, L.~Antiga, A.~Desmaison, A.~Kopf, E.~Yang, Z.~DeVito,
  M.~Raison, A.~Tejani, S.~Chilamkurthy, B.~Steiner, L.~Fang, J.~Bai, and
  S.~Chintala, ``Pytorch: An imperative style, high-performance deep learning
  library,'' in \emph{Advances in Neural Information Processing Systems},
  H.~Wallach, H.~Larochelle, A.~Beygelzimer, F.~d\textquotesingle
  Alch\'{e}-Buc, E.~Fox, and R.~Garnett, Eds., vol.~32.\hskip 1em plus 0.5em
  minus 0.4em\relax Curran Associates, Inc., 2019. [Online]. Available:
  \url{https://proceedings.neurips.cc/paper/2019/file/bdbca288fee7f92f2bfa9f7012727740-Paper.pdf}
\BIBentrySTDinterwordspacing

\bibitem{deepspeed-bert}
A.~D. authors, ``Deepspeed bert pre-training,'' 2022,
  \url{https://www.deepspeed.ai/tutorials/bert-pretraining/}.
  
\bibitem{megatron-lm}
S.~Mohammad, P.~Mostofa, P.~Raul, L.~Patrick, C.~Jared, and C.~Bryan,
  ``Megatron-lm: Training multi-billion parameter language models using model
  parallelism,'' \emph{arXiv preprint arXiv:1909.08053}, 2019.

\end{thebibliography}

\end{document}